# Biaxial growth of pentacene on rippled silica surfaces studied by rotating grazing incidence X-ray diffraction

Stefan Pachmajer [a], Oliver Werzer [b], Carlo Mennucci [c], Francesco Buatier de Mongeot [c], R. Resel [a,*]

(a) Institute of Solid State Physics, NAWI Graz, Graz University of Technology, Austria

(b) Institute of Pharmaceutical Sciences, Department of Pharmaceutical Technology, University of Graz, Austria

(c) Dipartimento di Fisica, Università di Genova, Via Dodecaneso 33, 16146 Genova, Italy

* Corresponding author.

E-mail address: roland.resel@tugraz.at (R. Resel)

**Abstract**

Pentacene is known to grow on isotropic silicon oxide surfaces in a substrate-induced phase with fiber textured crystallites. This growth study reports on the growth of pentacene crystallites on uniaxially oriented surfaces. Silica substrates have been treated by ion beam sputtering so that ripples with a lateral corrugation length of 38 nm and a surface roughness of 1.3 nm are formed. Pentacene thin films with a nominal thickness in the range from 20 nm up to 300 nm are deposited on top of the rippled surfaces. The films are characterized by atomic force microscopy and grazing incidence X-ray diffraction. Bi-axially oriented crystals are formed due to the grooves of the substrate surface opening up the possibility of a defined in-plane alignment of the crystals. In a first stage of thin film growth, the thin film phase (TFP) of pentacene is formed, while in the later stage the bulk crystal structure (C, Campbell phase) also appears. Due to the bi-axial alignment of the crystallites the transition from the thin film phase to the bulk crystal structure can be directly investigated. An epitaxial relationship with $(120)_{TFP} \parallel (210)_C$ and $[-210]_{TFP} \parallel [1-20]_C$ is observed which can be explained by an adaption of the herringbone layers of both crystal structures. This work reveals one possible microscopic mechanism for the transition from the metastable substrate-induced phase of pentacene to its equilibrium bulk structure.

1. **Introduction**

Polymorphism is a frequently discussed topic for pentacene within thin films[1]. Two polymorphic phases are found within macroscopic bulk crystals[2,3], but there are also two different substrate-induced phases which appear when substrates are present during a heterogenous crystallization process[4,5]. These substrate-induced phases are metastable and a transition to the thermodynamically stable bulk phase occurs via solvent vapor annealing or thermal treatment[6,7]. Thin film growth of pentacene on flat silica surfaces is dominated by the formation of the thin film phase directly at the substrate surface[8] where the molecules form crystals with the 001 plane parallel to the substrate surface. The main driving force of this preferred orientation is the formation of molecular layers parallel to the substrate surface. These layers are formed by up-right standing molecules which are packed in a herringbone arrangement[9–11]. Screw and edge dislocations are also reported as defects within pentacene films[12]. Above a critical film thickness a transition from the thin film phase to the Campbell phase has previously been observed and the microscopic mechanisms for this transition have been discussed in terms of crystal defects and self-limited growth[12,13].

The use of a substrate with defined surface corrugations – like the one formed on oxygen reconstructed Cu(110) surfaces – results in a different crystallization behavior of pentacene[14]. It is observed that the pentacene molecule conforms with the substrate so that the rod-like shape of the molecule fills the corrugated area of the substrate surface[15]. The subsequent crystal growth is bi-axially ordered with one defined crystallographic plane parallel to the substrate surface and with a defined azimuthal alignment of the crystals. However, unfavorable arrangements of the molecules in the first monolayer results in crystallization of pentacene with standing molecules and not lying as is often observed for other strongly interacting substrates[15]. The crystallization of pentacene on irregular or isotropic, rough substrates results in standing molecules being in contact with the surface[16]. It is suggested that the differences in the behavior on rough surfaces and flat surfaces results in higher heterogenous nucleation rates due to lower thermal desorption and moreover in reduced surface diffusion[17].

In this work, we study the crystallization of pentacene on an irregular corrugated surface consisting of uni-axially aligned surface ripples. A bi-axial growth of pentacene crystals is observed which allows a clarification of a possible microscopic mechanism for the transition from the thin film phase to the Campbell phase.

2. **Experimental**

Amorphous silicon dioxide substrates (Spectrosil 2000 by Heraeus) have been used as substrates. Ripples on the substrates have been prepared within a UHV chamber by ion beam treatment using $Ar^+$ ions accelerated with an energy of up to 800 eV towards the substrates up to a fluence of $1.2 \times 10^{18}$ ions/$cm^2$. An incident angle of 45° was chosen. The ripples formed by such a bombardment form perpendicular to the ion beam[18]. For the sake of comparability of all the samples and experiments, a sample coordination system is defined with the y-axis along the surface ripples, the x-axis perpendicular to the surface ripples and the z-axis along the substrate normal.

Pentacene, purchased from Sigma Aldrich and used without further purification, was deposited via physical vapor deposition. The Knudsen cell which was used for this process was mounted orthogonally to the substrate surface to avoid any possible shadowing effects from the surface ripples. Pentacene films with thicknesses of 20 nm, 40 nm and 300 nm were deposited using a growth rate of 0.5 nm/min, controlled via a quartz crystal microbalance. Substrates were kept at ambient temperature during all the different steps.

Atomic force microscopy (AFM) measurements have been performed on a Nanosurf Easyscan 2 with a 70 µm scan head in tapping mode. A silicon-spm-sensor tip of the type PPP-NCLR was used. Data were processed and analyzed using the software Gwyddion[19].

For the investigation of the crystallographic properties of the pentacene thin films various X-ray diffraction techniques were used to gain information on the alignment of the pentacene crystallites relative to the substrate surface and relative to the surface ripples. First, X-ray diffraction pole figures were performed using a lab-based Philips X-Pert system equipped with an Eulerian cradle. Radiation from a Chromium tube was used in combination with a secondary side graphite monochromator and a proportional counter as point detector. In the case of a pole figure, a constant diffraction angle (or scattering vector) was set and the angle $\varphi$ was varied continuously from 0° to 360° with intensity integration in steps of 3°, while the angle $\psi$ was varied in steps from 0° to 90 ° with a step size of 2°. It should be noted, that the weak diffraction of our thin film samples only allows measurement of the most intense reflections. An additional complication is that peaks at large $\psi$ – angles are difficult to detect[20]. The data evaluation was performed using the software STEREOPOLE[21].

Rotating grazing incidence X-ray diffraction (rot-GIXD) is a method that allows large volumes of reciprocal space to be measured from thin film samples. Further, the extension of the standard GIXD technique by an additional sample rotation enables the investigation of thin films with azimuthally (in-plane) aligned crystallites, i.e. uni-axially, bi-axially, or even fully epitaxially grown thin film crystallites. The measurements were performed at the beamline XRD1 at Elettra Sincrotrone (Trieste, Italy). An X-ray energy of 8.86 keV (wavelength of 1.40 Å) was used and diffracted intensity was collected by a Pilatus 2M detector (Dectris, Switzerland). For the measurements the sample was mounted on a setup utilizing a kappa-geometry. A defined incidence angle of 0.5° was used which is well above the critical angle of the pentacene films as well as that of the substrate material. For the rot-GIXD experiments, the samples were rotated azimuthally around the surface normal. An integration range of two degree was chosen, which means 180 images were collected during a full 360° rotation. Detector calibration and conversion of collected data to reciprocal space representation was done using the open access software program GIDVis[22]. The diffraction patterns are transferred to reciprocal space coordinates with defined directions of $q_x$ and $q_y$ pointing perpendicular to and along the ripples of the substrate, respectively; while $q_z$ points perpendicular to the substrate surface. $q_{xy}$ represents the in-plane part of the scattering vector by $q_{xy} = \sqrt{q_x^2 + q_y^2}$. Interpretation of the GIXD pattern was performed by calculated peak positions and peak intensities based on the known crystal structure solutions of pentacene[2,10]. The lattice parameters for this purpose are $a$ = 0.596 nm, $b$ = 0.760 nm, $c$ = 1.561 nm, $\alpha$ = 81.25°, $\beta$ = 86.56°, $\gamma$ = 89.80° for the thin film phase[10], while the lattice parameters $a$ = 0.790 nm, $b$ = 0.606 nm, $c$ = 1.601 nm, $\alpha$ = 101.9°, $\beta$ = 112.6°, $\gamma$ = 85.8° were used for the Campbell phase[2].

3. Results

*3.1. Substrate topography*

The bare substrate surfaces, i.e. without pentacene, have been investigated after ion bombardment by *ex-situ* AFM investigations. The substrate appears rather similar over the entire surface (see Figure 1A) and reveals two features of different length scale. First there are some grain-like structures with a separation of about 1µm; these are typical for the underlying silicon dioxide surface. More important for this study are the surface ripples which run from the bottom to the top of the image, i.e. along the y-direction. The characteristic ripple heights are determined to be in the range of 4 - 5 nm, but can even reach 10 nm. Calculation of the height-height correlation function allows the important parameters to be determined using the following expression[23]:

$$H(r) = 2\sigma^2 \left[1 - e^{-\left(\frac{r}{\xi}\right)^{2\alpha}}\right]$$

with σ being the root-mean-square roughness, ξ the lateral correlation length, and α the roughness exponent or Hurst parameter. The correlation function of the rippled substrate surface is depicted in Figure 2A. As expected from the anisotropic appearance in the AFM height images, the calculation reveals different height-height correlation functions when calculated in vertical (y – direction, black curve) and in horizontal (x – direction, red curve) directions. Starting with the evaluation of the slope at low relative distances r, the roughness exponent (0 ≤ α ≤ 1) is evaluated[24]. This parameter characterizes short range properties of the surface, with locally smooth surface structures resulting in high values of $\alpha$. A linear fit of the height-height correlation function in the low distance regime of Figure 2A reveals α = 0.7 for both directions which indicates a locally smooth surface. At large r, the height-height correlation function reaches a plateau proportional to $2\sigma^2$, from which the root-mean-square roughness calculates to $\sigma_{RMS}$ = 1.3 nm. The lateral correlation length, a characteristic or most common distance within two surface points with similar heights, is determined from the crossover region using linear extrapolation between the short range and long range regimes. This provides a value of $\xi_h$ = 38 nm for the x-direction and $\xi_v$ = 110 nm for the y-direction. This difference appears due to the surface anisotropy, i.e. the ripples running along the y-direction. Please note that there is a weak oscillatory behavior in the horizontal height-height correlation function which is absent in the other direction. These oscillations originate from the fact that the ripples are very similar over the surface, i.e. this reflects the periodicity of the ripple structure. It leads to a fourth parameter, the average ripple separation wavelength λ. From the current data, this is more easily extracted from the power-spectral-density function (data not shown) and gives a value of λ = 87 nm, which means that a typical peak – peak (or valley – valley) distance has this separation distance.

In a next step, the slopes across the ripples – given by the angle ß - are discussed. The density of the tilt angles ß is depicted in Figure 2B. The width of the angle distribution reveals that the surface inclines with respect to a perfectly flat surface only by few degrees. From the almost uniform distribution, one can conclude that the ripples have a rounded shape, where all angles between 0° and the maximum angles are present. The maximum angles are found to be 8.5° in either direction, i.e. areas incline up to this value relative to the average "mean" sample surface. Considering the central point of the broad distribution, it can be concluded that the ripple profile is rather symmetrical, in analogy to similar ion beam sputtering experiments[18,25]. This symmetry might appear surprising; since the ripples are formed perpendicular to the ion beam with a substrate tilt angle of 45°.

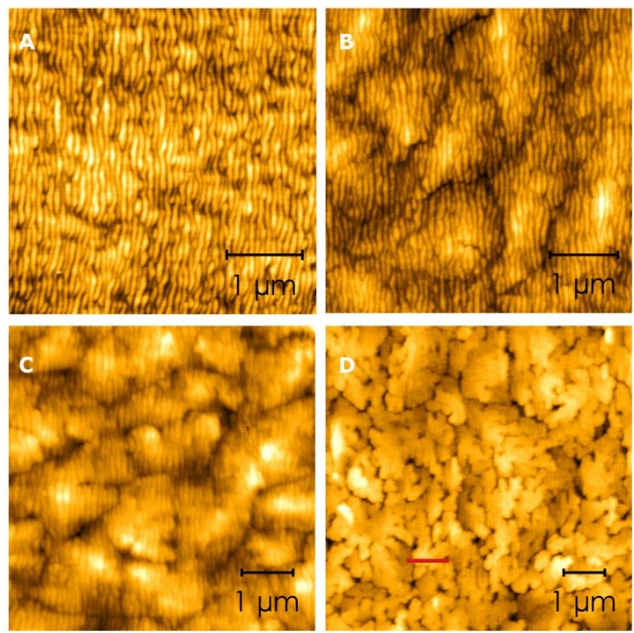

Figure 1: AFM topography images of pentacene thin films grown on a rippled silica surface. A) The bare substrate with ripples (z = 10 nm). B) rippled substrate with a nominal 20 nm thick pentacene film on top (z = 20 nm). C) a nominally 40 nm thick pentacene film (z = 40 nm). D) a nominally 300 nm thick pentacene film (z = 100 nm).

### 3.2. Pentacene topography on rippled surfaces

A pentacene thin film with a nominal thickness of 20 nm was deposited on the rippled substrate surface; Figure 1B depicts an AFM micrograph of this sample. Surprisingly, at first glance the image appears similar to the bare substrate (c.f. Fig. 1A), with the surface morphology revealing the ripples. The pentacene islands are hardly detectable, but compared to the bare substrate the root-mean-square roughness increased to $\sigma_{RMS}$ = 3.0 nm. This increase can be attributed to the growth of pentacene islands. This image also does not show any morphological anisotropy of the pentacene islands. This is different to other organic thin films which show an anisotropy when deposited on rippled surfaces[26].

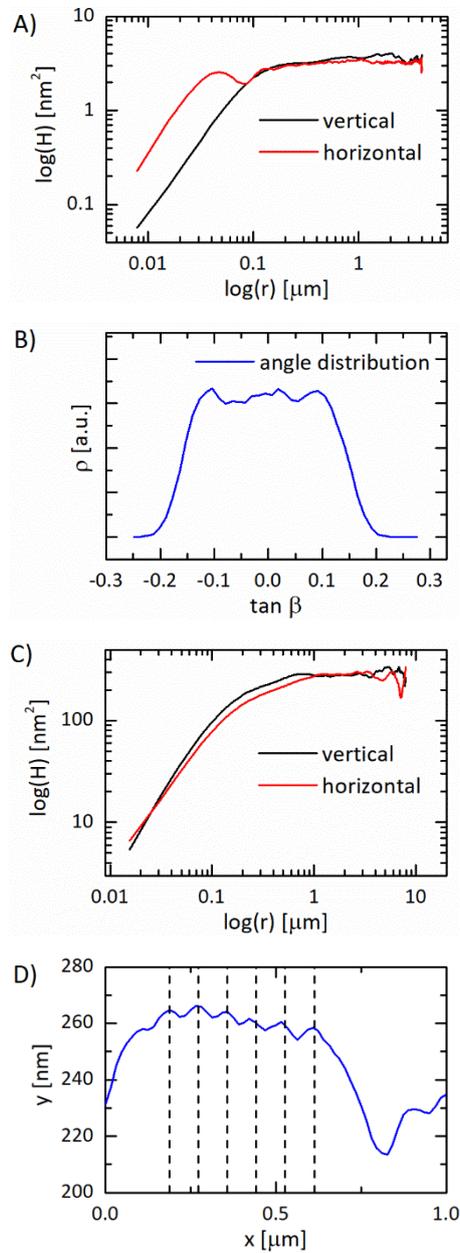

Figure 2: Analysis of the AFM investigations depicted in Figure 1A and 1D. A) Height-height correlation functions of the uncovered rippled surface in two different directions, B) distribution of the tilt angles across the surface ripples, C) height-height correlation functions of the 300 nm pentacene film in two different directions and D) height scan along a specific direction marked in Fig.1D (red line), the dotted lines represent a specific surface periodicity. The two different directions are across (horizontal or x-direction) and along (vertical or y-direction) the surface ripples.

The morphology of a pentacene film with a nominal thickness of 40 nm is shown in Figure 1C. The underlying ripples of the bare substrate are still visible, but in comparison to the 20 nm film, the pentacene islands are more easily distinguishable from the structure of the bare substrate. The shape of these islands is typical for pentacene when grown on flat surfaces. Besides this, an increase in the surface roughness to $\sigma_{RMS}$ = 5.2 nm is found which is larger than the thinner pentacene film or the bare substrate. Further, the bright spots in the AFM images are the top of the pyramidal islands which reflects the 3-D growth mode of pentacene which results from the effect known as self-roughening growth of pentacene[27].

Figure 1D depicts the morphology of a pentacene film with 300 nm nominal thickness, the corresponding height – height correlation function is given in Figure 2C. The surface roughness increases to $\sigma_{RMS}$ = 12 nm, i.e. the self-roughening continues. While the morphology appears similar to that of pentacene grown on flat surfaces, a closer inspection reveals still the presence of the ripple-like structures (line scan depicted in Fig 2D). Please note, that the visibility of these ripples within Fig.1D is limited in comparison to Fig.1A due to the contrast range, which has to span over a z-value of 100 nm in Fig.1D, while the same contrast range is available over only 10 nm in Fig.1A. Regular oscillations in the pentacene topography with an approximate peak-to-peak distance of 90 nm are observed. While those structures might also involve some terrace steps of pentacene, this distance is equal to the ripple separation wavelength of the bare rippled substrate. Also, the ripple height is in the same range as for the bare rippled substrate. This strongly suggests that the pentacene adapts to the underlying rippled structure and keeps memory of the underlying topography without leveling even at pentacene film thicknesses far from the substrate surface.

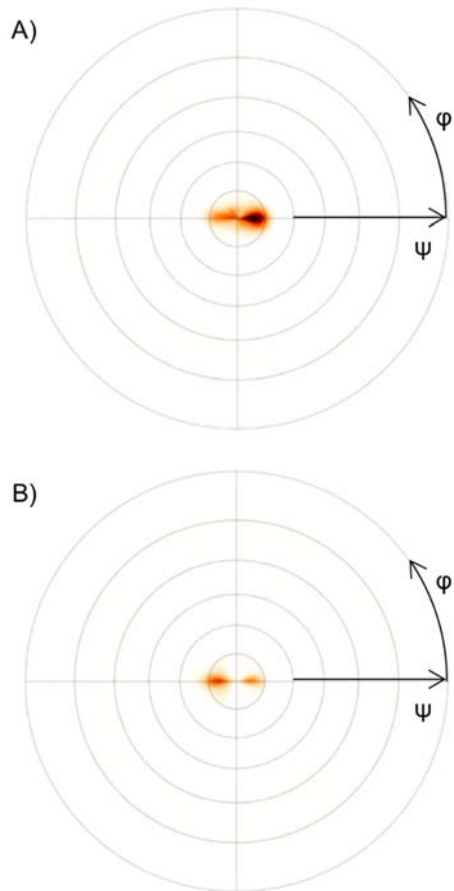

Figure 3: Pole figures of a 300 nm pentacene film for the 001 poles of the thin film phase (A) and pole figures for the 001 poles of the Campbell phase (B). The surface ripples are in vertical (y -) direction at $\varphi = 90°$.

### 3.3. Polefigure measurements

To determine the alignment of the pentacene crystals with respect to the ripples, the thin films were investigated by X-ray diffraction (XRD). Typically, for such experiments, one starts with a specular XRD scan to test which net plane of pentacene crystals is formed parallel to the substrate surface. Surprisingly, the specular XRD measurements did not reveal any diffraction peaks originating from the organic layer. This means that there is no crystallographic plane that is parallel to the substrate. By performing polefigure measurements, the orientation distribution in all directions above the sample horizon of known poles (or net plane normals) can be determined[28]. To begin with, the 300 nm sample is investigated; Figure 3A shows the polefigure for the 001 poles of the pentacene thin film phase. Two enhanced pole densities are observed within this experiment; both are tilted by 8° in the $\psi-$ direction away from the center of the polefigure (please note, intensity at the center of a polefigure would represent net planes which are parallel to the underlying substrate and provide the same information that would be observed in the specular scan.) From the location of the poles it follows that the 001 net planes are on average inclined by 8° to the "mean" surface. The azimuthal direction of these 8° tilted poles are along φ = 0° and 180° which coincides with the x-direction of the defined coordinate system. These directions are across the ripples of the substrate surface perpendicular to the ripples direction, so it can be concluded that the pentacene crystals align with respect to the ripple direction.

The second polefigure (Fig. 3B) gives the spatial distribution of the 001 poles of the Campbell phase, i.e. the bulk phase structure. Very surprisingly, the polefigure looks similar to that of the thin film phase; we observe two directions inclined by an angle of 8° away from the substrate surface normal with a tilt towards the ripples. If these results are compared with previous polefigure findings we can assess that the Campbell phase growth is impacted by the substrate surface in a similar manner to the thin film phase.

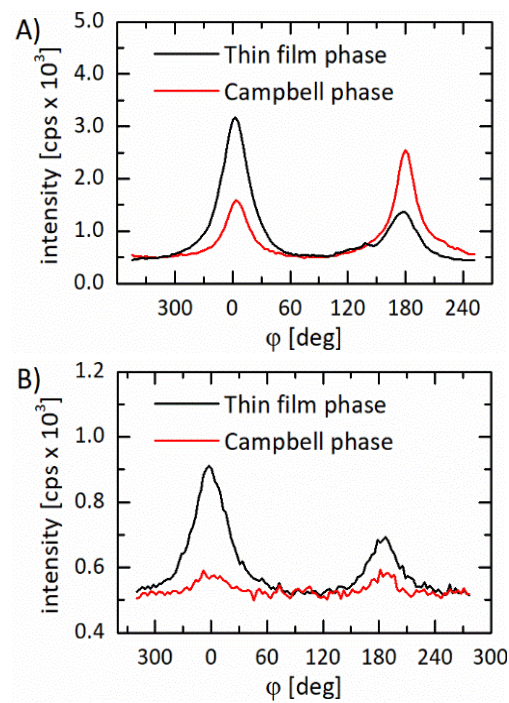

Figure 4: φ - scans at the 001 Bragg peak of the thin film phase and of the Campbell phase for the 300 nm (A) and for the 40 nm (B) thick pentacene films. The scans are performed at a constant angle ψ = 8°.

While the polefigures look similar, Figure 4A depicts details of both these pole figures to highlight their differences: φ scans at constant ψ = 8° for the 001 poles for the thin film phase and for the Campbell phase. The intensity distributions reveal two maxima at φ = 0° and φ = 180° for both of the phases which are the positions of the maximum pole densities within the polefigures in Fig. 3A and 3B.

Comparing the maxima of intensity for the two peaks of the thin film phase we observe that the observed intensity is different for the two different directions. This means that the thin film crystallites are more likely to form with an inclination of $\psi$ = 8° and $\varphi$ = 0°, with a minimum at $\varphi$ = 180°. The data for the Campbell phase shows that the majority of pentacene in the Campbell phase grows at $\varphi$ = 180°, with a minimum at $\varphi$ = 0°. Please note, that the dominant parts of the thin film phase and of the Campbell phase are in opposite directions, i.e. they are rotated by 180° with respect to each other. Comparing the intensities of both phases a relative percentage of 60% and 40% was estimated for the thin film phase and Campbell phase, respectively.

Repeating the same polefigure measurements for the 40 nm (data not shown) and extracting the same $\varphi$-scans at $\psi$ = 8°, a comparable picture with maxima at $\varphi$ = 0° and $\varphi$ = 180° is revealed (see Fig. 4B). The relative intensities of both peaks show that for the thin film phase again a majority of the crystallites appear with at a tilt of $\varphi$ = 0°. For the Campbell phase, the maxima position also remained the same, but the amount present for both directions is now very similar, i.e. their intensity ratio is nearly identical. One can conclude from this that, in the thin film of 40 nm, the development of a majority species in the Campbell phase preferring one direction is absent, whereby the thin film phase definitely keeps this preferred orientation. The comparison of the amount of thin film phase and Campbell phase shows that about 80% of pentacene deposited onto the surface is in the thin film phase while 20% is in Campbell phase. This is in good agreement with many observations showing that the thin film phase is dominant in the initial growth stages for pentacene crystals prepared on a flat surface; while the Campbell phase is dominant in thicker films[29]. The introduction of the ripples does not seem to disturb this behavior. While there is still some controversy about the initial development of the Campbell phase during deposition, i.e. if it is already present in the early growth stages[30], our results here show that the Campbell phase grows directly at the substrate surface, but to a lesser extent.

Summarizing the in-house pole figure investigations (Fig. 3 and Fig. 4), it is clear that the Campbell phase is weakly present in the 40 nm film and is becoming more dominant in the 300 nm film. Both polymorphs investigated here show a crystal alignment with the 001 poles (net plane normals) tilted 8° from the substrate surface normal in specific azimuthal directions. The majority of species alignments of both crystal types appear in opposite directions.

### 3.4. Rotating grazing incidence X-ray diffraction

To obtain further information on the in-plane orientation of pentacene crystals grown on rippled silicon oxide surfaces, grazing incidence X-ray diffraction with rotation of the samples was performed. While polefigures are necessary to access diffraction information close or equal to $\psi = 0°$, GIXD is particularly helpful for diffraction information present close to the in-plane direction ($\psi = 90°$) which is hardly accessible using a standard polefigure setup. In Figure 5A-C, reciprocal space maps for the nominally 300 nm thick film are shown, the reciprocal space maps are calculated from diffraction patterns taken with the primary X-ray beam aligned in 3 different azimuthal directions; two directions along the ripples and one direction across the ripples. Please note, the reciprocal space maps are plotted as a function of the in-plane part of the scattering vector $q_{xy}$, since a single diffraction pattern is not sufficient to calculate a reciprocal space map along a specific direction, e.g. x- or y-direction. Besides the diffraction patterns, calculated peak positions and peak intensities are also plotted. The thin film phase (black circles) as well as the Campbell phase (red circles) are provided using the previously determined information from the polefigure measurements, i.e. the 001 planes incline by 8° relative to the substrate surface. Please note, that the presented peak positions are based on the conclusion of ideally fibre textured crystallites being present. It means that the absence of a peak in the experiment would mean that this particular crystal orientation does not diffract at this specific sample position. For the sake of visibility, the Laue indices of the calculated peaks are only given in Figure 5D.

The discussion of the GIXD patterns starts around the specular direction close to $q_{xy} = 0$. In case of Fig. 5A as well as in Fig. 5B, one can see peaks from the 00L series of both the pentacene thin film phase and Campbell phase. In all cases the 00L peak series are tilted by about 8° from the surface normal. The presence of these peaks in a GIXD map is caused by the width of the 00L poles in $\psi$ - direction (compare Fig. 3A and 3B) as well as by the width of the 00L poles in $\varphi$ - direction (compare Fig. 4A). The rather broad width in $\psi$ - direction result in crescent shaped diffraction peaks in Fig. 5A and 5B, while the rather narrow width in the $\varphi$ – direction result in a strong variation of intensity of the observed 00L peaks. Considering the 8° tilt angle of the scattering vectors, it can be concluded that these observations are in perfect agreement to the pole figure measurements.

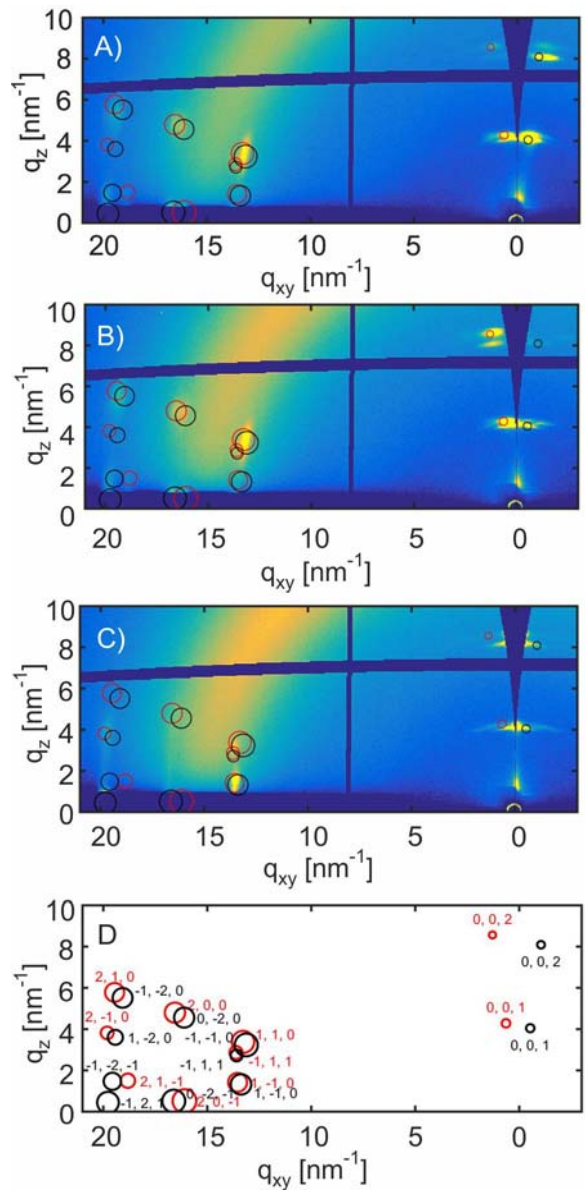

Figure 5: Reciprocal space maps for a nominally 300 nm thick pentacene film for different in-plane measurement directions. A) The primary X-ray beam along the substrate ripples ($q_{xy} \approx q_{-x} = q$ ($\varphi = 180°$)) with maximum intensity from 001 of 8° tilted thin film phase crystallites. B) The primary beam along the surface ripples in the opposite direction ($q_{xy} \approx q_x = q$ ($\varphi = 0°$)) with maximum intensity from 001 of 8° tilted Campbell phase crystallites. C) The primary X-ray beam across the ripples ($q_{xy} \approx q_y = q$ ($\varphi = 90°$)). D) Indexation with Laue indices for the thin film phase (black circles) and for the Campbell phase (red circles) choosing a texture with 8° tilt of the 001 pole directions relative to the substrate surface

In a following step, the diffraction peaks at larger scattering vectors (10 nm$^{-1}$ to 20 nm$^{-1}$) are analyzed. There are many different broad peaks meaning that nearly identical peak patterns are present in Fig.

5A and 5B. For peak indexation, it is difficult to assign the peaks unambiguously to either the thin film or Campbell phase. The identical peak positions at different azimuthal directions suggest an overlap of the in-plane peaks of the different phases present. The third diffraction pattern (Fig.5C) is qualitatively different with respect to the previous two patterns. Also here an unambiguous indexation of the diffraction peaks to one of the two phases cannot be performed, but the presence of different diffraction patterns at different azimuthal rotations indicates a certain in-plane orientation of the thin film crystallites.

Using the entire data set of a rotating GIXD experiment allows extraction of the complete information regarding the orientation of the two phases relative to the surface ripples. Polefigure representations from this data are found to be a useful tool where they can be calculated by selecting a certain q-value of interest and calculating the corresponding $\varphi$ and $\psi$ values for each data point with the selected q. Details on this calculation can be found within the open access software program GIDVis[22]. Figure 6A shows a pole figure calculated at the position q = 19.8 nm$^{-1}$ to detect the thin film phase. The poles of 120 are arranged along a ring with a maximum located at $\varphi$ = 180°. Additionally, the -120 poles are visible as well in this pole figure. Please note, that this pole figure separates crystal orientations with parallel 001 and 00-1 poles (or (001) and (00-1) contact planes) which cannot be distinguished by the 001 pole figures (Fig. 3A). The related pole directions of 120 and -120 are indicated by different symbols using crosses or circles. The strong smearing along the rings reveals a strong preferred orientation of crystallites, but with a rather weak in-plane order which might be labeled as a large in-plane mosaicity. From this polefigure analysis, it can be determined that the real space axis or [-210] direction of the thin film phase ([-210]$_{TF}$) is parallel to the surface ripples (y - direction).

A second pole figure was calculated at 20.3 nm$^{-1}$ to analyze the Campbell phase crystallites based on its 210 and 2-10 poles. The 210 pole of the Campbell phase is clearly present at $\varphi$ = 180°, while the 2-10 peak is expected to be less pronounced due to its considerably smaller structure factor. The smearing of the poles along $\varphi$ reveals, even in this case, weak in-plane order of the crystallites. The crystallographic [1-20] direction of the Campbell phase ([1-20]$_C$) is aligned parallel to the surface ripples (y-direction).

Combining the results of both pole figures allow to conclude on the relation of the thin film phase with the Campbell phase so that an epitaxial relationship between the dominant fractions of the two phases can be given. The equal direction of the 120 pole of the thin film phase and the 210 pole of the Campbell phase (compare Fig. 6A and 6B) reveal that the (120) plane of the thin film phase is parallel to the (210) plane of the Campbell phase. Also specific crystal directions are related: $[-210]_{TF}$ is parallel to $[1-20]_C$; both are parallel to the surface ripples. The degree of in-plane orientation is comparable for both types of phases. It is only weakly pronounced, which can be easily seen by the smeared – crescent shaped – distribution of pole intensities. A calculation of an orientation function is difficult to perform due to overlap of diffraction peaks with different Miller indices. However, an in-plane mosaicity spread of about 60° can be given for both types of crystals.

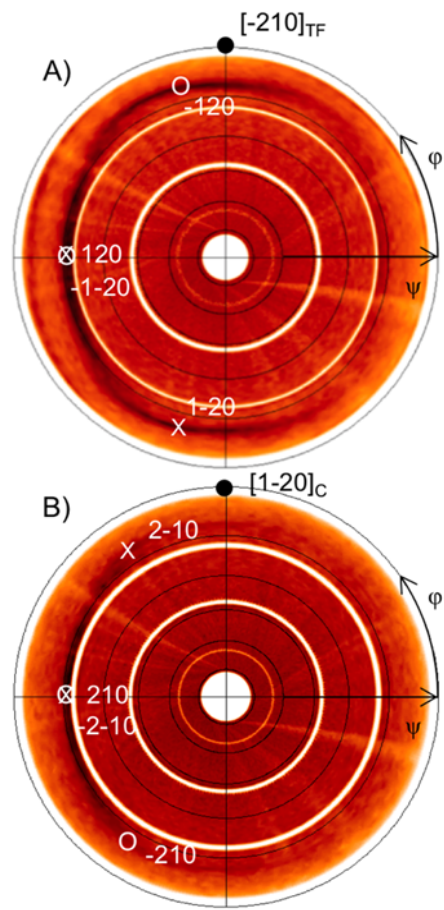

Figure 6: Pole figures calculated from rotating GIXD experiments at q = 19.8 nm$^{-1}$ detecting the 120 and -120 poles of the thin film phase (A) and at q =20.3 nm$^{-1}$ detecting the 210 and 2-10 poles of the Campbell phase (B). Crosses and circles give the pole directions for the two equivalent contact planes (see text). The crystallographic directions which are parallel to the surface ripples are given by black dots.

Knowing of the crystal alignments from the pole figures, we can determine in a subsequent step how the molecules are aligned relative to the surface ripples and how the molecules of the two phases are oriented relative to each other. Figure 7 shows the alignment of the molecules along the surface ripples. For simple visualization, we draw the (120) plane of the thin film phase as well as the (210) plane of the Campbell phase. In both cases these two planes run along the surface ripples (Fig. 7A) and a zig-zag arrangement of neighboring molecules is observed along the surface ripples (Fig. 7B).

An interesting result is the relation of the molecular packing of both phases relative to each other. Identical features for both crystal phases are observed: i) the long molecular axes of the pentacene molecule have the same alignment and ii) the aromatic planes enclose the same tilt angle with respect to neighboring molecules. An adaption of the herringbone layer of the thin film phase to that of the Campbell phase is geometrically possible. The (120) plane of the thin film phase and the (210) plane of the Campbell phase act as a quasi-twinning plane for connection of the Campbell phase with the thin film phase. This twinning plane leads to a defined angle between the (001) planes of both phases (compare Fig.7) relative to each other. Both planes enclose an angle of 16° which explains the 8° tilt of both phases in either direction quite well.

4. Discussion

The deposition of pentacene on flat $SiO_x$ surfaces typically results in the formation of the thin film phase already in the initial growth stages. On the deposition of more material, the Campbell phase become evident and develops more significantly compared to the thin film phase as the thickness increases. A very similar situation is observed for the growth on a rippled surface, a small amount of Campbell phase is present in the 40 nm film, but becomes more dominant in thicker films of 300 nm. This means that the thin film phase becomes less dominant with increasing film thickness. Recent results suggest the extinction of the thin film phase is a result of self-limited growth so that at larger film thickness the Campbell phase is favored[13].

The growth of pentacene islands on flat $SiO_x$ surfaces also results in the formation of islands which after the first monolayers develop a strong tendency towards 3D island formation with distinct terraces of about 1.5 nm step height[30]. This step height is prototypical for pentacene molecules growing in an upright standing configuration. However, in contrast to the behavior on flat surfaces, the morphology in our case is overlaid by a different morphology. On a small scale (high resolution), this overlay results from the surface ripples which dominate the morphology, so that the characteristic terraced structure cannot be observed in the films with thicknesses of 20 nm, 40 nm, or even at 300 nm, where the rippled surface is already fully overgrown by pentacene crystallites. While it was shown that the flexibility of aromatic crystals can allow the islands to adapt for the change in the local morphology[31], in our case it cannot be determined if there is a bending of the crystals or if the morphology results from introduction of stacking adaptions/faults on account of this local variation of the substrate height.

The growth of pentacene reveals large differences in the diffraction pattern when either a flat surface or a rippled surface are used as the substrate. Specular X-ray diffraction on films grown on flat surfaces reveals a typical fiber texture with a common (001) contact plane with the substrate surface, but a random in-plane (azimuthal) behavior[8,32]. In our case here of rippled surfaces, no diffraction peak appears in the specular X-ray diffraction data at all. In fact, an 8° tilt of the 001 planes of the thin film phase, as well as of the 001 planes of the Campbell phase, is found as determined by the polefigure and the rot-GIXD measurements.

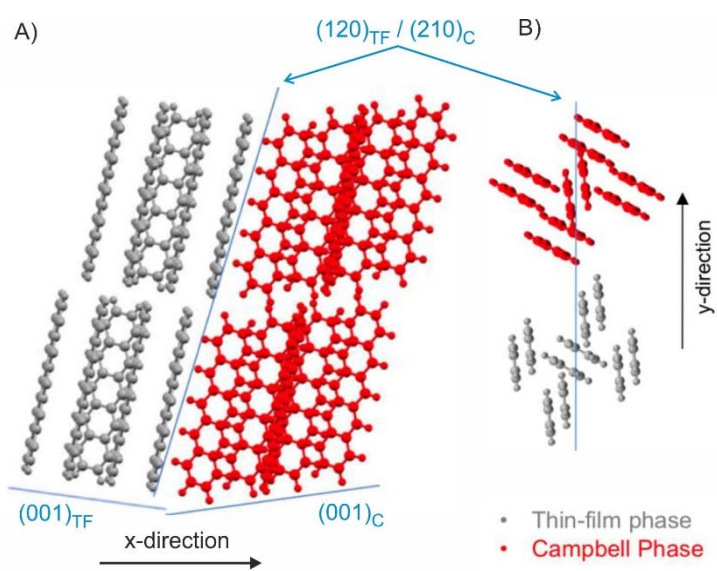

Figure 7: Orientation of the two pentacene phases – the Campbell phase (C) and the thin film phase (TF) - relative to the surface ripples (indicated by x and y - direction) and relative to each other: A) in a side view along the surface ripples and B) in a top view. Crystallographic planes are drawn by blue lines and denoted by their Miller indices.

It is generally accepted that the formation of a substrate-induced phase (like the pentacene thin film phase) is associated with a confinement of the molecular packing due to a flat substrate surface[33,34]. The nucleation at step edges is not observed for pentacene on silicon dioxide surfaces; therefore, it is surprising that the thin film phase is formed on rippled surfaces. Obviously, a substrate surface with a roughness of 1.8 nm, a roughness exponent of 0.7 and correlation lengths of 38 nm and 110 nm provides sufficient flat areas, so that a substrate-induced phase can be formed. The question arises why an 8° inclination of the crystallites is formed most often. The identified maximum surface slopes of the substrate at 8.5° might correlate to the nucleation sites of the crystallites due to sufficiently extended flat areas. Such a flat area of maximum tilt might coincide with an inflection point about half way down the ripple as schematically drawn in Figure 8. Due to such areas, this might also explain the crystal growth along this area, parallel to the ripple direction. From this the bi-axial alignment, i.e. the in-plane direction, might also be explained.

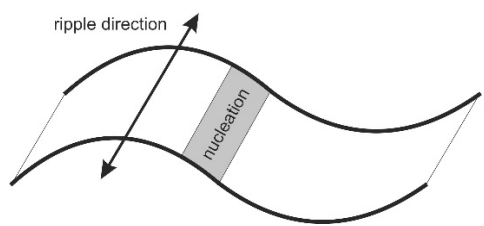

Figure 8: Schematic picture of surface ripples with the nucleation area for pentacene crystallites along the inflection points of the surface ripples.

The experiments reveal, that there is a preferred orientation of the crystal with an 8° slope. While the surface appears rather symmetric from microscopic investigations, the asymmetric implantation process could cause small differences in the nature of the surface on either side of the ripple. This difference (e.g. in surface energy or nanoscopic surface roughness) might result in a side with favorable adsorption, nucleation and crystal growth conditions compared to the other side. However, as the crystals overgrow the entire ripples, the morphology does not show any signs of asymmetric growth.

In the in-plane direction, the dominating crystal growth directions of the thin film phase and of the Campbell phase are analyzed in combination with each other (compare Fig.7). It turns out that the packing of the molecules along the herringbone layers are comparable for these two growth directions: i) nearly identical molecular packing within the herringbone layers and ii) similar alignment of the molecules relative to the substrate surface.  As a consequence, different alignments of the herringbone layers (or of the (001) planes) relative to the substrate surface are observed. It results in the (001) planes or the two phases enclosing a tilt angle of 16° with each other. We suggest that the Campbell phase nucleates directly at the (120) plane of the thin film phase by adaption of the herringbone packing of the molecules.

## 5. Conclusion

This work investigates the growth of pentacene crystals on a rough silicon oxide surface. A substrate with uniaxially aligned surface ripples is chosen with a ripple height of a few nm and with an average lateral periodicity of about 87 nm. In the first growth stage the thin film phase of pentacene is formed, but also a small fraction of Campbell phase crystallites appears. At the advanced thin film growth stage, Campbell phase crystallites becomes more dominant. Biaxially aligned crystals are formed with a highly defined out-of-plane alignment of the crystallites, but with weak in-plane alignment. The dominant part of the pentacene thin film phase crystallites are analyzed in relation to the dominant part of the Campbell phase crystallites. A transition from the metastable thin film phase to the thermodynamically stable Campbell phase is concluded by the nucleation of the (210) plane of the Campbell phase at the (120) plane of the thin film phase. An explanation is given at the molecular scale by an adaption of the herringbone packing by the molecules of both phases.

**Acknowledgements**

Financial support was given by the Austrian Science Foundation (FWF): [P25887]. The authors thank the synchrotron ELETTRA, Trieste, Italy for providing synchrotron radiation and allocation of beamtime at the beamline XRD1.